\documentclass[reprint,amssymb,amsmath,aps,prb,showpacs]{revtex4-1}
\usepackage[dvips]{graphicx}
\usepackage{bm}
\usepackage{color}
\begin{document}
\title{Giant Positive Magnetoresistance and field-induced metal insulator transition in Cr$_2$NiGa}
\author{S. Pramanick$^{a}$} 
\author{P. Dutta$^{b}$}
\author{S. Chatterjee$^{b}$}
\author{S. Giri$^{a}$}
\author{S. Majumdar$^{a*}$}
\affiliation{$^a$Department of Solid State Physics, Indian Association for the Cultivation of Science, 2A \& B Raja S. C. Mullick Road, Jadavpur, Kolkata 700 032, India}
\affiliation{$^{b}$UGC-DAE Consortium for Scientific Research, Kolkata Centre, Sector III, LB-8, Salt Lake, Kolkata 700 098, INDIA} 
\email{sspsm2@iacs.res.in}

\begin{abstract}
We report here the magneto-transport properties of the newly synthesized Heusler compound  Cr$_2$NiGa which  crystallizes in a disordered cubic B2 structure belonging to {\it Pm$\bar{3}$m} space group. The sample is found to be paramagnetic down to 2 K with metallic character.  On application of magnetic field, a significantly large increase in resistivity is observed which corresponds to magnetoresistance as high as 112\% at 150 kOe of field at the lowest temperature. Most remarkably, the sample shows negative temperature coefficient of resistivity below about 50 K under the application of field $\geq$ 80 kOe, signifying a field-induced metal to `insulating' transition. The observed magnetoresistance follows Kohler's rule below 20 K indicating the validity of the semiclassical model of electronic transport in metal with a single relaxation time. A multi-band model for electronic transport, originally proposed for semimetals, is found to be appropriate to describe the magneto-transport behavior of the sample.   
\end{abstract}

\maketitle

\section{Introduction}
Intermetallic Heusler ~\cite{heusler} system of alloys continue to be the forefront of active research  since its discovery about a century ago.  Heusler alloys reveal wealth of fascinating  properties including  half-metallicity~\cite{hase,miura,jourdan}, magnetic shape memory effect~\cite{kainuma,krenke}, unconventional superconductivity~\cite{sc} and so on. In general, they represent a class of ternary intermetallic compounds having general formula X$_2$YZ where X and Y are transition metal whereas Z is a nonmagnetic {\it sp} element. Ideally, Heusler compounds crystallize in the Cu$_2$MnAl-type ordered L2$_1$ structure consisting of eight stacked body-centered-cubic lattices. The most common disordered structure arising from L2$_1$  is B2 (CsCl-type structure) which occurs due to random occupancy of Y and Z atoms. In this case Y and Z sites become equivalent with the lowering of crystal symmetry (space group no. 221: {\it Pm$\bar{3}$m})~\cite{graf}.

\begin{figure}[t]
\centering
\includegraphics[width = 7.5 cm]{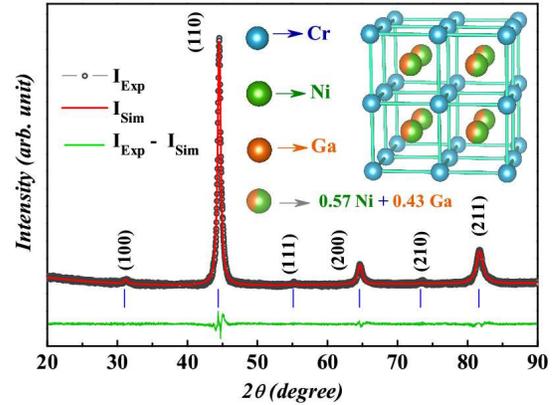}
\caption {Room temperature XRD of pattern of Cr$_2$NiGa (data points) recorded on a Bruker AXS diffractometer (D8 advance) using Cu K$_{\alpha}$ radiation along with the Rietveld refinement curves (solid line) .The structure illustrates the 3D-network in disordered cubic Heusler phase of Cr$_2$NiGa having disorder between Ni and Ga.}
\label{xrd}
\end{figure}
\par
Recently, Cr-based stoichiometic (Cr$_2$YZ) full Heusler compounds have become an interesting topic of investigations. Particularly, systems having 24 valance electrons are  promising candidates for being half metallic compounds~\cite{sasi,mein,haki}. Among them, Cr$_2$CoGa (crystallizes in the XA inverse Heusler structure) has been predicted to be a half-metallic fully compensated ferrimagnet  where the moments from two Cr sites (at X and Z positions) compensate each other resulting zero macroscopic moment.~\cite{sasi} We have here studied the nearest sibling of the compound namely, Cr$_2$NiGa which has 25 valance electrons. Till now there is no report of the formation of the compound  experimentally nor there is any report on the electronic or magnetic properties based on theoretical calculations. In contrast to the Co-counterpart, Cr$_2$NiGa is found to be paramagnetic (PM) with rather unusual magneto-transport properties at low temperature.     

\section{Experimental Details}
Polycrystalline sample of Cr$_2$NiGa was prepared by standard arc melting technique. Next, the arc molten ingot was vacuum sealed in a quartz capsule and annealed at 900$^{\circ}$ C for 48~h followed by normal furnace cooling down to room temperature.  In order to get confirmation about the  stoichiometry of the alloy, we performed energy-dispersive x-ray spectroscopy (EDS) by scanning different parts  (effective scanning area of 0.5$\times$0.5 mm$^2$) of the sample using JEOL JSM 6700F FESEM. The average atomic stoichiometry was found to be Cr = 1.984, Ni = 1.044 and Ga = 0.972, which are close to the expected  values. The crystallographic structure was investigated by X-ray powder diffraction (XRD) and the Rietveld refinement was performed using MAUD program package~\cite{maud}. Magnetic measurements were carried out by using a commercial Quantum Design SQUID magnetometer (MPMS XL Ever Cool model). The resistivity ($\rho$) was measured by four probe method on a homemade setup fitted in a nitrogen cryostat (between 77 and 400 K) as well as on a cryogen-free high magnetic field system (Cryogenic Ltd., U.K.) between 5-300 K.  

\section{Results}
\subsection{Powder X-ray Diffraction}
Fig.~\ref{xrd} depicts powder XRD pattern of the studied sample measured at room temperature. Observation of some strong Bragg reflections in the XRD pattern initially indicate a simple body centered cubic (bcc) structure. Rietveld refinement of the pattern confirms the formation of CsCl-type B2 structure, where  Ni and Ga atoms randomly occupy Y and Z sites. We can simply rule out the formation of ordered L2$_1$ or XA structure from the absence of superlattice line around 2$\theta \sim$ 26$^{\circ}$. Best fitting is obtained by considering partial occupancies of Ni and Ga for the equivalent Y/Z position of X$_2$YZ Heusler structure with standard deviation $\sigma$ = 1.5046.  For the refinement, Ni occupancy was 0.57 whereas Ga has 0.43 occupancy and calculated cubic lattice parameter is found to be 2.8967 \AA.

\subsection{Magnetization}
\begin{figure}[t]
\centering
\includegraphics[width = 7.8 cm]{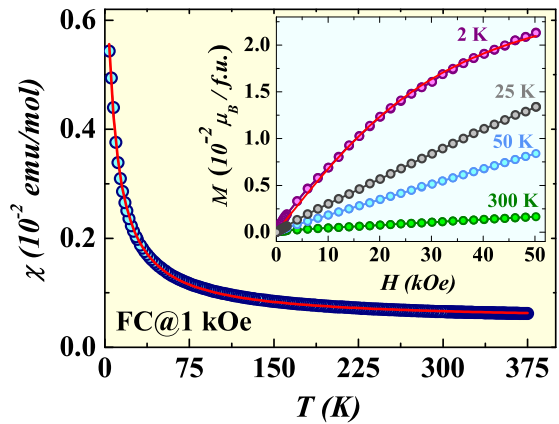}
\caption {Main panel shows $\chi$ {\it vs.} $T$ curve for $H$ = 1 kOe during cooling. Inset shows isothermal $M$-$H$ curves measured at different constant temperatures. Red solid lines are the fitted curves to the experimental data.}
\label{mag}
\end{figure}

We  presented the  temperature ($T$) variation of magnetic susceptibility ($\chi = M/H$, where $M$ is the measured magnatization and $H$ is the applied magnetic field) in presence of $H$ = 1 kOe from 380 to 2 K in fig.~\ref{mag}.  It is evident that $\chi$ increases monotonically with decreasing $T$ and we do not observe any signature of magnetic anomaly. The $\chi(T)$ data can be well fitted with the Curie-Weiss law (solid line in the plot) where an additional temperature independent susceptibility term is considered ($\chi = C/(T-\theta) + \chi_0$). The effective PM moment per formula unit and the PM Curie temperature  of the sample are  found to be $\mu_{eff}$ = 0.71 $\mu_B$, and  $\theta$ = -6.4 K respectively, while the value of $\chi_0$  ($\sim$ 10$^{-5}$ emu/mol) is found to be small but positive. The presence of small but finite $\chi_0$ is possibly due to the spin fluctuation enhanced Pauli paramagnetism in this itinerant electron system~\cite{goto}. 
\par
The PM nature of the  sample is further clarified from the isothermal magnetization measurement. The inset of  fig.~\ref{mag}  represents $M(H)$ curves measured at $T$ = 2, 25, 50 and 300 K for 0 to 50 kOe field cycling. The 2 K isotherm shows clear signature of curvature, while the other iostherms are found to be linear in nature.  We argue that the curvature is the manifestation of  inherent non-linear nature of the $M-H$ curves for free PM moments ($\mu$) which is particularly visible at low $T$ and high $H$ ($k_BT/\mu H \ll$ 1). We used classical Langevin function for PM material to fit the data with $M(x) = M_0[\coth(x) - 1/x]$, with $x$ = $\mu H/k_BT$. The fitting is shown in the inset of  fig.~\ref{mag} by solid line for 2 K isotherm. The good fit once again indicates the PM nature of the solid. Evidently, the sample has a quite low value of $\mu$ (0.0214 B $\mu_B$/ f.u. at 2 K for $H$ = 50 kOe), which is presumably related to the itinerant nature of the Cr or Ni moments. 

\subsection{Magneto Transport}
\begin{figure}[t]
\centering
\includegraphics[width = 7.5 cm]{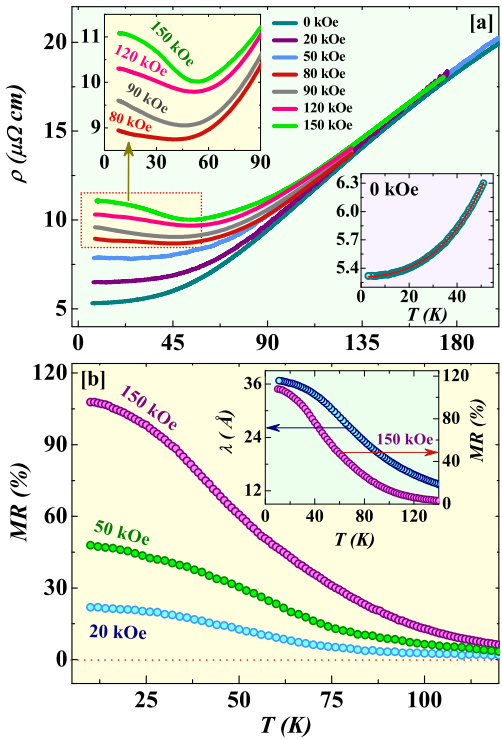}
\caption {Main panel of (a) shows $\rho$ measured during heating in presence of different $H$. Upper inset shows enlarged view of low-$T$ part. Lower inset depicts fitting of the data with  $\rho(T) = a_0 + a_2T^2 + a_4T^4$ below 50 K. (b) shows $T$ variation of iso-field MR for different constant $H$. Inset shows the variation of $\lambda$ and $\Delta \rho/\rho_0$ (for 150 kOe) with temperature.}
\label{rho}
\end{figure}

The most fascinating observation in the present work is obtained from the magneto-transport measurement on the sample. The  $T$ variation of $\rho$ during heating in presence of $H$ = 0, 20, 50, 80, 90, 120 and 150 kOe are shown in the fig.~\ref{rho}(a). For all the measurements, the magnetic field was in transverse direction with respect to the current. The zero-field $\rho(T)$ curve shows pure metallic nature as evident from continuous drop in $\rho$ with decrease in $T$. The residual resistivity ratio (RRR = $\rho$(300 K)/$\rho$ (4 K)) is found to be 5 with the value of $\rho$ (4 K) = 5 $\mu\Omega$-cm, signifying good metallic character of the sample. All the  $\rho(T)$ curves measured in field overlapped with the $H$ = 0 curve in the high $T$ region (above about 135 K), whereas, deviation between them is visible at low temperatures. $\rho(T)$ becomes flat below about 50 K for $H$ = 20 and 50 kOe. However, for $H$ = 80 kOe and above, $\rho(T)$ show an upward turn around 50 K and increase ($d\rho$/$dt <$ 0) down to the lowest temperature. Such upturn indicates that the system undergoes a field-driven transition from a pure metallic to a state having `insulating' character. For clear visualization, low $T$ parts of $\rho$ for $H$ = 80, 90, 120 and 150 kOe are shown in the upper inset of fig.~\ref{rho}(a). Large increment in $\rho$ with $H$ at low $T$ signifies large positive magnetoresistance (MR) in the sample. Thermal variations of calculated MR $\left[= (\rho_H - \rho_0)/\rho_0= \Delta \rho/\rho_0\right]$ from the different iso-field ($H$ = 20, 50 and 150 kOe) $\rho(T)$ curves are plotted in the main panel of fig.~\ref{rho}(b). Here $\rho_0$ and $\rho_H$ are resistivity in zero field and in presence of $H$. We observe a huge 112\% of positive MR at 8 K in $H$ = 150 kOe. The MR is also reasonably high at lower $H$, for example, MR is about 50\% at low temperature ($\sim$ 7 K) for $H$ = 50 kOe. With increasing $T$, magnitude of MR decreases and tends to vanish as $T$ approaches 135 K.

\par
\begin{figure}[b]
\centering
\includegraphics[width = 7.5 cm]{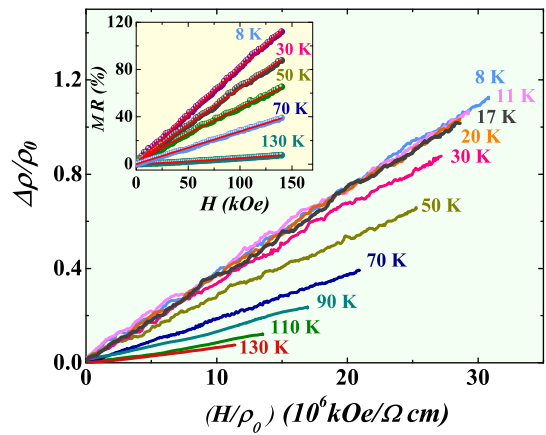}
\caption {Main panel of represents the Kohler's plot of the measured isotherms. Inset shows isothermal $H$ variation of MR measured at selective temperatures. Red lines are the fitted curves to the isotherms.}
\label{mr}
\end{figure}

We have carried out isothermal $H$ variation of $\rho$ at different constant $T$. Since significant field effect is only observed below 135 K, we have restricted our measurement up to this temperature. For this purpose, $\rho$ was measured at selected constant $T$ by cycling $H$ between $\pm$ 140 kOe and the corresponding MR was calculated (see inset of fig.~\ref{mr}). For the sake of clarity, only positive $H$ quadrant is presented. Apparently, all the measured isotherms show quasi-linear nature with varying slope depending upon $T$. The 8 K isotherm shows MR of magnitude 112 \%, which is in accordance with the iso-field $\rho(T)$ data. At high-$T$, MR is quite small and for example, it is about 8\% at  130 K.
\par
Positive MR in a metal is often attributed to the curving of electron trajectory due to the application of $H$ which effectively reduces the electronic mean free path ($\lambda$).~\cite{pippard,ziman} This model gives positive MR only when there are  more than one type of carriers and a quadratic dependence of MR on $H$ is obtained. For the present sample, MR is quasi-linear contrary to the usual observation. We have fitted the field dependence of MR data by the relation $\Delta \rho/\rho_0$ = $\alpha H^n$ where $\alpha$ is the coefficient of power law. The values of $n$ are found to be close to unity and lie between 0.89 to 1.23 which is well below 2 expected for a PM or non-magnetic metal.

\par
We have carefully looked at the low-$T$ part of the data by plotting $\rho$ as a function of $T$. $\rho$ does not obey a simple $T^2$ dependence in accordance with the Fermi liquid behavior. However, there are numerous reports in the literature, where the low-$T$ part of a Fermi metal can be described by a slightly modified law containing $T^4$ term resulting $\rho(T) = a_0 + a_2T^2 + a_4T^4$. The fitted curve is shown by the solid line in the lower inset of  fig.~\ref{rho} (a). The values of the coefficient  ($a_0$ = 5.3 $\times$ 10$^{-6}$ $\mu\Omega$-cm, $a_2$= 2.22 $\times$ 10$^{-10}$ $\mu\Omega$-cm$T^{-2}$ and $a_4$ = 6.21 $\times$ 10$^{-14}$ $\mu\Omega$-cm$T^{-4}$) are at par with the reported data for other metallic magnetic systems.~\cite{kaul}     

\par
MR in a metal is often found to follow a scaling law, known as Kohler's rule. If $\Delta \rho$ is the change in $\rho$ due to the application of  $H$, then a conventional metal obeying Fermi liquid behavior can follow an empirical relation : $\Delta \rho/\rho_0$ = $\mathcal{F}(H/\rho_0)$, where $\mathcal{F}$ is a function characterized by  the inherent electronic structure and external geometry of the metal.~\cite{pippard,luo}  Here $\rho_0$ can be  expressed as  $\rho_0 = m^{*}/ne^2\tau$ for free electrons with $\tau$ being the characteristic relaxation rate of scattering. $m^{*}$, $n$ and $e$ are the effective electronic mass, volume density of conduction electrons and electronic charge respectively. In the light of semiclassical theory of transport,  Kohler's rule should only be valid  for a metal having single type of charge carrier along with the equal $\tau$ at all the points on the Fermi surface. Accordingly, $\Delta \rho/\rho_0$ plotted as a function of $\mathcal{F}(H/\rho_0)$ should be $T$ independent. Since our data indicates quasi-linear variation of MR with $H$, we have chosen $\mathcal{F}(H/\rho_0)$ to be equal to  $H/\rho_0$. Interestingly, all the MR curves in the Kohler's plot up to 20 K collapse on each other signifying the validity of the rule. However, above 20 K we see marked deviation from the Kohler's rule as the curves are found to be strongly $T$ dependent. Such violation of Kohler's rule at high-$T$ is likely to be related to the enhanced electron-phonon scattering which breaks the uniqueness of $\tau$.             

\par
We have calculated  $\lambda$ of the conduction electrons on the framework of Drude-Sommerfeld model where 
$$\lambda = \tau v_F = \frac{\hbar}{ne^2\rho}\left(3\pi^2n\right)^{1/3}$$ Here $v_F$ is the Fermi velocity.~\cite{ash}
The value of $\lambda$ is found to be around 40 \AA~ at 5 K, which indicates that $\lambda$ is large enough (it is much larger than the interatomic spacing) so that the system can be described by the semiclassical model. We have plotted thermal variation of $\Delta \rho/\rho_0$ ($H$ = 150 kOe)   along with the calculated $\lambda(T,H=0)$ in the inset of fig.~\ref{rho} (b). Interestingly, thermal variation curves for both the parameters are quite similar signifying close correspondence between mean free path of scattering (and as a result $\tau$) and the magnitude of MR. Such correspondence as well as the validity of Kohler's law (at least for $T \leq$ 20 K)  imply that the observed magneto-transport behavior of Cr$_2$NiGa can be well accommodated within the semiclassical model for electronic conduction in metals with a single $\tau$.~\cite{ziman,abrikosov}

\section{Discussion}
The observed MR in Cr$_2$NiGa is substantially high particularly for a polycrystalline intermatellic sample with static defects and disorders (residual resistivity is about 5 $\mu \Omega$-cm). The present sample does not show any signature of long range magnetic ordering down to 2 K. Therefore, the observed large positive MR is unlikely to have a magnetic origin. Such conclusion is also supported by the fact that our MR data corroborate well with the free electron model. Even if there is some magnetic correlation present in the sample, it is not playing any role towards the magneto-transport properties of the system.  

\par
There are few examples of large positive MR in bulk intermetallic systems,~\cite{chandan,roop,juan} and in many cases, MR is found to be uncorrelated with the ordered magnetic state of the system. The most important observation in the present study is the remarkable metal to `insulator' transition in high applied fields. Possibly, the trend for a semiconducting nature starts at a lower field (the positive MR is an outcome of that), eventually leading to a state with negative $T$ coefficient of $\rho$. The semiconducting bahaviour of the sample can have several possible origins, such as weak localization, Kondo effect or due to the opening up of a soft gap at the Fermi level. We can rule out localization or Kondo process as in general $H$ should  drive the system to have lower $\rho$ causing negative MR. The $\rho$ versus $T$ plot under $H$ = 150 kOe also do not follow $\log T$ variation below the $\rho$ minimum. We also failed to observe any activated behavior  ($ \rho \sim \exp{(-1/T)^{\beta}}$, $\beta$ is a generalized parameter $\leq$ 1) that fits well with an electronic gap. A gap in the Fermi level would also expect to alter $m^{*}/ne^2$, which in its turn should violate the Kohler's rule~\cite{luo}.

\par
Interestingly, similar positive MR and metal-insulator transition had been reported for several semimetals~\cite{ali,yuan,shekhar} including Bi, Sb and graphite~\cite{du,li} (reported mostly on high mobility single crystalline samples). Surprisingly, the magneto-transport properties of these semimetals qualitatively resemble quite well with the presently studied Cr$_2$NiGa compound. In case of semimetals mentioned above, the insulating state develops at a rather low magnetic field (can be as low as 1 kOe)~\cite{shekhar} as compared to hundred of  kOe in the present Heusler system. Nevertheless, the field induced `insulating' part shows similar $T$ variation with very identical saturating behavior at the lowest $T$. Among many plausible explanations, a simple multi-band  model has been found to be quite successful in explaining such metal to insulator transition as well as positive MR. In this model, total $\rho(T)$ is found to contain a $H$ independent metallic part and a $H$ dependent `insulating' part which vanishes for $H$ = 0~\cite{shekhar}. The `insulating' part dominates at lower $T$ surpassing the metallic contribution and thereby turning overall $\rho$ to increase with decreasing $T$. Although, semi metallicity in full Heusler alloys are not known, similar multi-band model can be generalized and found useful for explaining these anomalous magneto-transport properties of the studied compound.                  
 

\end{document}